\documentclass[12pt,a4paper]{article}
\pdfoutput=1
\usepackage{jheppub}
\usepackage[table]{xcolor}

\usepackage[T1]{fontenc}
\usepackage[utf8]{inputenc}
\usepackage{amsmath}
\usepackage{amssymb}
\usepackage{esint}
\usepackage{microtype} 
\usepackage{lmodern} 
\usepackage{booktabs}
\usepackage{multirow}
\usepackage{mdframed}
\usepackage{graphicx}
\usepackage{amsfonts}
\usepackage{bbm}
\usepackage{amssymb}

\usepackage{slashed}
\usepackage{young}
\usepackage[vcentermath]{youngtab}
\usepackage{tensor}
\usepackage{ulem}

\usepackage{pdflscape}

\providecommand*{\dd}{\mathop{}\!d}
\renewcommand*{\dd}{\mathop{}\!d}

\providecommand*{\pd}{\mathop{}\!\partial}
\renewcommand*{\pd}{\mathop{}\!\partial}

\providecommand*{\hs}{\mathop{}\!\star}
\renewcommand*{\hs}{\mathop{}\!\star}

\newcommand{\cF}{\mathcal{F}}

\usepackage{mathtools}
\newcommand{\bslashed}[1]{\mathrlap{#1}\backslash}

\title{Three-dimensional conformal geometry and prepotentials for four-dimensional fermionic higher-spin fields}

\author[a,1,2]{Marc Henneaux\note{On leave of absence from Coll\`ege de France, Paris}\note{\href{https://orcid.org/0000-0002-8912-6384}{ORCID: 0000-0002-8912-6384}},}
\author[a,b,3]{Victor Lekeu\note{\href{https://orcid.org/0000-0001-6111-005X}{ORCID: 0000-0001-6111-005X}},}
\author[a,c]{Amaury Leonard,}
\author[a,4]{Javier Matulich\note{\href{ https://orcid.org/0000-0002-3558-9025}{ORCID: 0000-0002-3558-9025}},}
\author[a,5]{and Stefan Prohazka\note{\href{https://orcid.org/0000-0002-3925-3983}{ORCID: 0000-0002-3925-3983}}}

\affiliation[a]{Universit\'e Libre de Bruxelles and International Solvay Institutes, 
  Campus Plaine---CP~231, \\
  B-1050  Bruxelles, Belgium}
\affiliation[b]{The Blackett Laboratory, Imperial College London,\\ Prince Consort Road London SW7 2AZ, U.K.}
\affiliation[c]{Max-Planck-Institut f\"{u}r Gravitationsphysik (Albert-Einstein-Institut),
Am M\"{u}hlenberg 1, \\ DE-14476 Potsdam, Germany}

\emailAdd{henneaux@ulb.ac.be}
\emailAdd{vlekeu@ulb.ac.be}
\emailAdd{amaury.leonard@ulb.ac.be}
\emailAdd{jmatulic@ulb.ac.be}
\emailAdd{stefan.prohazka@ulb.ac.be} 

\abstract{We introduce prepotentials for fermionic higher-spin gauge
  fields in four spacetime dimensions, generalizing earlier work on
  bosonic fields. To that end, we first develop tools for handling
  conformal fermionic higher-spin gauge fields in three dimensions.
  This is necessary because the prepotentials turn out to be
  three-dimensional fields that enjoy both ``higher-spin
  diffeomorphism'' and ``higher-spin Weyl'' gauge symmetries. We
  discuss a number of the key properties of the relevant Cotton
  tensors. The reformulation of the equations of motion as ``twisted
  self-duality conditions'' is then exhibited. We show next how the
  Hamiltonian constraints can be explicitly solved in terms of
  appropriate prepotentials and show that the action takes then the
  same remarkable form for all spins.}

\begin{document}

\maketitle

\section{Introduction}
\label{sec:introduction}

Our paper is devoted to fermionic higher-spin conformal geometry in
three dimensions and its application to the study of the dynamics of
fermionic higher-spin gauge fields in four spacetime dimensions. More
precisely, we consider massless fermionic higher-spin gauge fields of
spin $s + \frac12$, described by tensor-spinors
$\psi_{i_1 \cdots i_s}$ that are totally symmetric in their $s$
indices $i_1, \cdots, i_s$. Under (linearized) higher-spin
diffeomorphisms and (linearized) higher-spin Weyl transformations,
these fields transform as
\begin{equation}
  \Gamma \psi_{i_1 i_2 \cdots i_s}
  = s \partial_{(i_1} \xi_{i_2 \cdots i_s)}
  +  s \gamma_{(i_1} \lambda_{i_2 \cdots i_s)} \, ,
  \label{Transf1a}
\end{equation}
where $\xi_{i_1 \cdots i_{s-1}}$ and $\lambda_{i_1 \cdots i_{s-1}}$
are symmetric tensor-spinors with $s-1$ indices. The first part of
this transformation is also called a generalized diffeomorphism and
the second one a generalized conformal transformation. The bosonic
analogs of these transformations for a spin-$s$ field
$Z_{i_1 i_2 \cdots i_s}$ are
\begin{equation}
  \Gamma Z_{i_1 i_2 \cdots i_s}
  = s  \partial_{(i_1} \xi_{i_2 \cdots i_s)}
  + \frac{s (s-1)}{2} \delta_{(i_1 i_2} \lambda_{i_3 \cdots i_s)} \, ,\label{Transf1aBose}
\end{equation}
where $\xi_{i_1 \cdots i_{s-1}}$ and $\lambda_{i_1 \cdots i_{s-2}}$
are now symmetric tensors with $s-1$ indices and $s-2$ indices (and no
spinor index), respectively. We have assumed for definiteness that the
metric is Euclidean and given by $\delta_{ij}$, as this is the case
relevant below. In the Minkowskian case, one simply needs to replace
$\delta_{ij}$ by the Lorentzian metric $\eta_{ij}$ in
\eqref{Transf1aBose}. A central question investigated here is the
construction of the invariants under both higher-spin diffeomorphisms
and higher-spin Weyl transformations and the study of their
properties. This is what we mean by ``developing conformal geometry''.
 
Conformal higher-spin gauge fields have attracted a lot of attention
over the years, in any number of spacetime dimensions (for more
information, see for instance some of the earliest references
\cite{Fradkin:1985am,Pope:1989vj,Fradkin:1989md,Fradkin:1989xt,Vasiliev:2001zy,Segal:2002gd,Shaynkman:2004vu,Vasiliev:2007yc,Metsaev:2007rw,Metsaev:2009ym}
and references therein). Conformal higher-spin gauge fields are
interesting per se, but also appear as ``prepotentials'' in manifestly
duality-invariant formulations of higher-spin {\em non-conformal}
gauge fields in 4 spacetime dimensions, through the resolution of the
constraints appearing in the Hamiltonian formalism.\footnote{The term
  ``prepotential'' is always used here in that sense, as potentials
  needed to solve the constraints of the Hamiltonian formalism.}
Following the work of \cite{Deser:1976iy}, this was originally
observed for spin-$2$ in
\cite{Henneaux:2004jw,Bunster:2012km,Julia:2005ze} and generalized to
higher integer spins in \cite{Henneaux:2015cda,Henneaux:2016zlu}. The
prepotentials appear as three-dimensional tensors defined on the
constant time Euclidean hypersurfaces in the 3+1 Hamiltonian spacetime
split. That they enjoy the higher-spin conformal gauge symmetry
\eqref{Transf1aBose} was somehow unexpected but established for all
integer spins \cite{Henneaux:2015cda,Henneaux:2016zlu} using the
relevant higher-spin conformal techniques in three dimensions.

For fermionic fields, however, prepotentials with the desired
properties were introduced only for spins $\frac32$
\cite{Bunster:2012jp} and $\frac52$ \cite{Bunster:2014fca}, where they
were also verified to enjoy the symmetries \eqref{Transf1a}. It was
conjectured in that latter reference that a similar pattern would also
hold for half-integer spins equal to $\frac72$ or higher, but that
conjecture was not proven.

One reason that the conjecture was left unproved in
\cite{Bunster:2014fca} is that the corresponding tools for handling
the higher-spin conformal symmetry in the dimension three relevant for the
construction of prepotentials were not available in a form adapted to
the Hamiltonian constraint analysis. The difficulty with dimension three
is that conformal symmetry is not controlled by the Weyl tensor, which
identically vanishes, but by the Cotton tensor, which involves higher
derivatives of the fields.

The Cotton tensor has been defined for higher-spin bosonic gauge
fields in \cite{Damour:1987vm}, \cite{Pope:1989vj}, and
\cite{Henneaux:2015cda}\footnote{Note a small subtlety between
  \cite{Pope:1989vj} and \cite{Henneaux:2015cda} when $s \geq 4$. It
  is that the definition given in \cite{Pope:1989vj} involves a
  symmetrization by hand, which turns out not to be necessary because
  the relevant expression is actually symmetric. This observation
  turns out to be useful for establishing the properties of the Cotton
  tensor.} and its properties relevant to the introduction of
prepotentials through the resolution of the Hamiltonian constraints
have been established in \cite{Henneaux:2015cda} (see also
\cite{Basile:2017mqc}). The Cotton tensor contains $s-1$ derivatives
of the Riemann tensor and thus $2s-1$ derivatives of the higher-spin
$s$ bosonic field. It plays a central role both in the study of the
dynamics of conformal higher-spin gauge fields in three spacetime
dimensions
\cite{Bergshoeff:2009tb,Bergshoeff:2011pm,Nilsson:2013tva,Nilsson:2015pua,Linander:2016brv}
and, as we have just pointed out, for the introduction of
prepotentials in the Hamiltonian formulation of standard higher-spin
gauge fields in four spacetime dimensions.

The purpose of this article is to extend the work of
\cite{Henneaux:2015cda,Henneaux:2016zlu} to higher-spin fermionic
fields. To that end, we define and study the properties of the Cotton
tensor for half-integer spin fields in three dimensions. The Cotton
tensor is actually a tensor-spinor, but like for any other
tensor-spinor we shall often loosely refer to it just as tensor. The
Cotton tensor contains $2s$ derivatives of the field
$\psi_{i_1 i_2 \cdots i_s}$ (in terms of the spin $S = s + \frac12$,
this is equal to $2S -1$ as in the bosonic case). It was defined
earlier in \cite{Andringa:2009yc} for spin $\frac32$ and more recently
for all half-integer spins in
\cite{Kuzenko:2016bnv,Kuzenko:2016qdw,Kuzenko:2016qwo}. Our derivation
follows a different line. It is based on the use of the differential
operator $d_{(s)}$ of
\cite{DuboisViolette:1999rd,DuboisViolette:2001jk} that fulfills
\begin{equation}
d_{(s)}^{s+1} = 0
\end{equation}
and the corresponding Poincaré-type lemmas. As such, our definition it
is not tied to supersymmetry or superspace calculus. The same method
has been applied to mixed Young symmetry tensors for which the ``critical dimension'' where the Weyl tensor identically vanishes is
generically higher than 3 (see, e.g., \cite{Bunster:2013oaa} and
\cite{Henneaux:2016opm,Henneaux:2017xsb,Lekeu:2018kul,Henneaux:2018rub}).

Once the Cotton tensor has been defined and its main properties
established, one can turn to the resolution of the fermionic
constraints of the Hamiltonian formalism. These can be rewritten in a
form that makes the introduction of prepotentials effortless.

Our paper is organized as follows. Section \ref{Cotton} is devoted to
the definition and study of the properties of the Cotton tensor for
half-integer spin fields in three dimensions. We then consider in
Section \ref{Hamiltonian} the dynamics. We first show that the
equations of motion can be rewritten, just as in the bosonic case
\cite{Henneaux:2016zlu}, as twisted self-duality conditions
(\cite{Deser:1977ur,Deser:2004xt,Deser:2014ssa}). We then turn to the
Hamiltonian formulation of the equations of motion, in particular to
the constraint equation, which plays a central role in the twisted
self-duality conditions. We solve in Section \ref{Prepotentials} the
constraints, which is the step that introduces the prepotentials in
terms of which we rewrite the action. This action enjoys a chiral
$SO(2)$ symmetry. Section \ref{Conclusions} is devoted to final
comments and conclusions.
Table \ref{tab:summary}, appended at the end of this work, summarizes the most important definitions and properties of bosonic
and fermionic higher spin fields in the prepotential formalism and
might be useful to get a fast overview.

\paragraph{Notation and conventions.}

The flat metric of $4$-dimensional spacetime has signature
$(-,+, +,+)$ and its spatial sections are Euclidean with signature
$(+,+,+)$. Our convention for the Dirac $\gamma$ matrices is that
$\{\gamma_{\mu},\gamma_{\nu}\} = 2 \eta_{\mu\nu}$ where
$\eta_{\mu\nu}$ is the spacetime metric. Furthermore, we define
$\gamma_5 \equiv \gamma_0 \gamma_1 \gamma_2 \gamma_3$ , so that the
spatial gamma matrices satisfy the useful identity
\begin{equation}
  \label{eq:usefid}
\gamma_i \gamma_j = \delta_{ij} + \varepsilon_{ijk} \gamma^k\gamma_5 \gamma_0  \, ,
\end{equation}
and
$\gamma_{ij} \equiv \gamma_{[i} \gamma_{j]} = \varepsilon_{ijk}
\gamma^k\gamma_5 \gamma_0$ with
$\varepsilon_{123} = \varepsilon^{123} = 1$. Notice that
$(\gamma_{5})^{2}=-I$.

Taking one spatial trace is indicated with a bar,
$\bar{T} =T^{[1]}= \delta^{ij} T_{ij}$, and the slash is the spatial
gamma-trace, $\slashed{T} = \gamma^i T_i$. Multiple traces are
indicated by a bracketed exponent, e.g.,
$T^{[2]} = \delta^{ij} \delta^{kl} T_{ijkl}$. Due to the property
$\gamma^{(i} \gamma^{j)} = \delta^{ij}$, the double gamma-trace of a
symmetric tensor is the same as a normal trace, so no notation is
introduced for multiple gamma-traces. Space-time traces are rarely
used and indicated by a prime or backslash respectively,
$T' = \eta^{\mu\nu} T_{\mu\nu}$ and $\bslashed{T} = \gamma^\mu T_\mu$.
The Dirac conjugate $\bar{\psi} = \psi^\dagger \gamma^0$ is also
indicated by a bar, but no confusion should arise since the context is
clear.

We will sometimes find it convenient to adopt a compact notation where
the vectorial indices are suppressed and symmetrization over
unwritten vectorial indices is implied. For example, in this notation, equations 
\eqref{Transf1a} and \eqref{Transf1aBose} become
\begin{align}
  \Gamma \psi &= s  \partial \xi  +  s  \gamma \lambda  \, , & \Gamma Z &= s \pd \xi + \frac{s(s-1)}{2}\, \delta \lambda \, .
\end{align}
Since it should be clear from the context and to improve readability we mostly call tensor-spinors just tensors.

Finally, a tensor(-spinor) with $(a_{1},a_{2},\cdots,a_{n})$ Young symmetry is
labeled by the length of the rows, i.e., corresponds to a Young
diagram with $n$ rows which have $a_{i}$ boxes. If not stated
otherwise, we follow the manifestly antisymmetric convention.

\section{Three-dimensional conformal geometry}
\label{Cotton}

The Riemann tensor, or equivalently the Einstein tensor, controls
higher-spin diffeomorphisms. By this we mean that any function that is
higher-spin diffeomorphism invariant can be written as a function of
the Riemann (or equivalently Einstein) tensor and its derivatives.
However, the Riemann tensor lacks higher-spin conformal invariance,
which is an important property needed for the resolution of the
Hamiltonian constraints. For this reason the Cotton tensor must be
introduced and its important properties established.

\subsection{Riemann tensor}

The ``Riemann'', or ``curvature'' tensor is defined by taking $s$
derivatives of the spin-$s$ (or $s + \frac12$) field
\cite{deWit:1979sib}. In terms of the differential operator $d_{(s)}$
of \cite{DuboisViolette:1999rd,DuboisViolette:2001jk}, the Riemann
tensor can be written as
\begin{equation}
  \label{eq:riemanndef}
  R = d_{(s)}^{s} \psi
\end{equation}
or, in components,
\begin{equation}
R_{i_1 j_1 \cdots i_s j_s} = 2^s \pd_{[j_1|} \cdots \pd_{[j_s|} \psi_{|i_1] \cdots |i_s]} \, .
\end{equation}
It is a tensor of Young symmetry type $(s,s)$ which satisfies the
Bianchi identity $d_{(s)} R =0$ because of the property $d_{(s)}^{s+1} = 0$. On
account of that same equation, it is also invariant under higher-spin
diffeomorphisms, which can be written as
$\Gamma_\xi \psi = d_{(s)} \xi$. Furthermore, a necessary and
sufficient condition for the higher-spin field to be a pure
higher-spin diffeomorphism, i.e., $\psi = d_{(s)} \xi$ for some $\xi$,
is that its Riemann tensor vanishes. This is equivalent to the
statement that the most general higher-spin diffeomorphism invariant
function can be expressed as a function of the Riemann tensor and its
derivatives only.

The Riemann tensor is not invariant under higher-spin Weyl
transformations. The construction of invariants for that symmetry
makes dimension three very special. In dimension strictly greater than
three, one can construct invariants by removing gamma-trace terms from
the Riemann tensor, defining thereby the Weyl tensor. This procedure
does not yield quantities of great interest in dimension three, however,
because the tracefree part of the Riemann tensor then identically
vanishes. What controls higher-spin Weyl symmetry is the Cotton
tensor, which contains higher derivatives of the higher-spin fields,
to which we will turn after defining the Einstein tensor for general
$s$.

\subsection{Einstein tensor}

As we indicated, the Weyl tensor identically vanishes in three
dimensions. The curvature tensor is therefore completely determined by
the ``Ricci tensor'', or equivalently, by the ``Einstein tensor'',
which is the $s$ times dual (with a sign factor inserted for convenience)
\begin{equation}
G = (-1)^s \, \underbrace{\hs \hs \cdots \hs}_{\hbox{$s$ times}} \, d_{(s)}^s \psi  
\end{equation}
of the curvature. We dualize on each antisymmetric pair so this
expression can also be written as
\begin{equation}
  G = \left(\varepsilon \cdot \partial \, \cdot \,\right)^s  \psi  \, .
\end{equation}
In words, the Einstein tensor is obtained by contracting $s$ times
$ \varepsilon_{ijk_r} \partial^j$ with
$\psi^{k_1 \cdots k_s}$.
Explicitly, when $s=1$ (spin $\frac32$) one has
\begin{equation}
  G_{i} = \varepsilon_{ijk}  \partial^j \psi^{k} \, ,
\end{equation}
while for $s=2$ (spin $\frac52$)
\begin{equation}
  G_{ij} = \varepsilon_{ikm} \varepsilon_{jln} \partial^k \partial^l \psi^{mn} \, .
\end{equation}
The Einstein tensor is a completely symmetric tensor which
fulfills the contracted Bianchi identity
\begin{equation}
\partial_{i_1} G^{i_1 i_2 \cdots i_s} = 0 \label{BianchiE0} \, .
\end{equation}
Conversely, any symmetric and divergenceless tensor can be written as
the Einstein tensor of some field.

While equivalent, we find it convenient in the sequel to work
systematically with the Einstein tensor rather than with the Riemann
tensor.

\subsection{Schouten and Cotton tensors: first cases}
\label{sec:cotton-tensor}

The Einstein tensor and its derivatives provide a complete set of
higher-spin diffeomorphism invariant functions, but little can be said
about higher-spin Weyl symmetry without introducing the Cotton tensor.
The idea is to algebraically construct out of the Einstein tensor and
its successive traces the ``Schouten tensor'' that transforms under
higher-spin Weyl transformations into a symmetrized gradient. The
Cotton tensor is then the Einstein tensor of the Schouten tensor and
is therefore Weyl invariant. This is much along the lines of the
bosonic case, as can be seen in Table~\ref{tab:summary} at the end of this work,
but there the Schouten tensor transforms under a
symmetrized double gradient rather than a single symmetrized gradient.
The symmetrized double gradient is removed by acting with
$d_{(s)}^{s-1}$ rather than with $d_{(s)}^s$, which explains the
difference in the number of derivatives when expressed in terms
of~$s$.

After the Cotton tensor is properly defined, we prove the following two
important theorems.
\begin{description}
\item[Gauge completeness:] The Cotton tensor is zero if and only if the
  field is pure gauge with respect to higher-spin diffeomorphisms and
  higher-spin Weyl transformations.

  This property ensures that the Cotton tensor fully controls the
  gauge invariance, which means that any local higher-spin diffeomorphism
  and higher-spin  Weyl invariant function can be written in terms of the Cotton tensor and
  its derivatives. So the Cotton tensor and its derivatives provide a
  complete set of gauge invariant functions (see, e.g., Appendix B.1
  of \cite{Henneaux:2015cda} for more details and the proof that this
  claim is equivalent to the property above).
  
\item[Conformal Poincaré lemma:] Any symmetric, divergenceless and
  gamma-traceless tensor can be written as the Cotton tensor of some field.
  
  This property will be crucial when solving the Hamiltonian
  constraint of the higher-spin fermionic field in Section
  \ref{Prepotentials}.
\end{description}

As a warm-up, we start with the spin-$\frac{3}{2}$ field. Most of the
subtleties of the general case are already present in the case of the
spin-$\frac{5}{2}$ field, which we discuss next before we generalize
to general spin.

\subsubsection{\texorpdfstring{Spin-$\frac32$}{Spin-3/2}}

We first consider the familiar case of spin $\frac32$ ($s=1$). The
spin-$\frac 32$ field is a vector-spinor $\psi_{i}$. We are looking
for a complete set of functions of this field invariant under the
following transformations
\begin{equation}
\Gamma \psi_{i} = \partial_{i} \xi + \gamma_{i} \lambda \, . \label{Spin32}
\end{equation}
As we have recalled, a complete set of invariants under spin-$\frac32$
diffeomorphisms, i.e., the first term in \eqref{Spin32}, is given by
the Einstein tensor
\begin{equation}
G_{i} = \varepsilon_{ijk}  \partial^j \psi^{k}  
\end{equation}
and its derivatives,
which transforms under a conformal transformation as
$\Gamma G_{i} = \varepsilon_{ijk}  \partial^j \gamma^k \lambda$.
This implies
$\Gamma \slashed{G} = -  2 \gamma_5 \gamma_0 \slashed{\partial} \lambda$
and leads us to define a Schouten tensor as
  \begin{equation} \label{eq:schouten32}
S_{i} = G_{i}  -  \frac{1}{2}  \gamma_{i} \slashed{G} \,.
\end{equation}
Its gauge variation is indeed a gradient
\begin{equation}
  \Gamma S_{i}
  = 
  \partial_{i} \left(\gamma_5 \gamma_0  \lambda \right) \,.
\end{equation}
The definition \eqref{eq:schouten32} is invertible: the Einstein
tensor can be expressed in terms of the Schouten as
\begin{equation}
  \label{G_S1}
  G_{i} = S_{i}  - \gamma_{i} \slashed{S} \, .
\end{equation}
Since the Einstein tensor is identically divergenceless, the Schouten satisfies
\begin{equation}
  \label{BianchiS1}
  0 = \partial^i S_{i}  -  \slashed{\partial} \slashed{S} \,. 
\end{equation}
This is the Bianchi identity for the Schouten tensor.
The conformal invariant is then the Einstein tensor of the
Schouten tensor, which we name ``Cotton tensor''. For the
spin-$\frac32$ field this quantity is given by
\begin{align}
D_{i} &=
\varepsilon_{ijk}  \partial^j S^{k}  \label{Cottino1}
  \\
      &= \frac{1}{2} \left(\partial_i \partial^j \psi_j  -  \Delta \psi_i \right)
         -  \frac{1}{2}  \varepsilon_{ijk}  \gamma_5 \gamma_0\slashed{\partial} \partial^j \psi^k  \, .  \label{Cottino}
\end{align}
It is identically divergenceless, and also gamma-traceless on account
of \eqref{BianchiS1} (and \eqref{eq:usefid}). It is called ``Cottino''
in \cite{Andringa:2009yc} where it was first introduced.

We will now prove that the Cotton tensor and
its derivatives provide a complete set of invariant functions with
respect to higher-spin diffeomorphism and Weyl transformations (gauge completeness),
and any tensor that is both gamma-traceless and divergenceless is the
Cotton tensor of some vector-spinor (``conformal Poincaré lemma'').

\paragraph{Gauge completeness.}
\label{sec:gauge32}

The first property is equivalent to the fact that $D_i = 0$ is a
necessary and sufficient condition for the spin-$\frac32$ field to be
pure gauge. By construction, we have $\Gamma D_{i}=0$ which shows that
for a pure gauge field the Cotton tensor necessarily vanishes. This
condition is also sufficient, since if
$D_{i} =\varepsilon_{ijk} \partial^j S^{k}= 0$, then (using the
Poincaré lemma with a spectator spinor index) we have
$S_{i} = \partial_{i} \rho$ for some $\rho$ that we can always write
as $\rho = \gamma_5 \gamma_0 \lambda$. Inserting now $S_{i}$ into
\eqref{G_S1} leads to
$ G_{i} - \varepsilon_{ijk} \partial^j \gamma^k \lambda =0$, or
equivalently
$\varepsilon_{ijk} \partial^j (\psi^{k} - \gamma^k \lambda )=0$. We
can again use the Poincaré lemma, yielding
$\psi_{i} = \partial_{i} \xi + \gamma_{i} \lambda$ for some $\xi$.
Therefore a vanishing Cotton tensor also implies that the field is
pure gauge.

\paragraph{Conformal Poincaré lemma.}

Furthermore, by running backwards the construction of the Cotton
tensor, it is also easy to see that any vector-spinor field $T_{i}$
that is both gamma-traceless and divergenceless, $\partial^i T_{i}= 0$
and $ \slashed{T}=0$, is the Cotton tensor of some vector-spinor field
$\psi_j$, i.e., $T=D[\psi]$.

Indeed, the condition $\partial^i T_{i}= 0$ implies
$T_{i} =\varepsilon_{ijk} \partial^j S^{k}$ for some $S^{k}$ that
fulfills $\partial^i S_{i} - \slashed{\partial} \slashed{S} = 0$ on
account of $\slashed{T}=0$. We can now define a tensor $G_i$ through
$G_{i} = S_{i} - \gamma_{i} \slashed{S}$: it fulfills
$\partial^i G_{i}= 0$ and is thus itself equal to
$G_{i} =\varepsilon_{ijk} \partial^j \psi^{k}$ for some $\psi_i$ which
is the searched-for vector-spinor.

\subsubsection{\texorpdfstring{Spin-$\frac52$}{Spin-5/2}}
\label{sec:sp52}

We now turn to the discussion of the spin-$\frac52$ field, a symmetric
tensor(-spinor) $\psi_{ij}$, which will lay the ground work for the
next section where general half-integer spin is considered. We are
looking for a complete set of functions that is invariant under the
transformations
\begin{equation}
\Gamma \psi_{ij} = 2  \partial_{(i} \xi_{j)}  +  2  \gamma_{(i} \lambda_{j)} \, .
\end{equation}
As we have seen, such a complete set of invariants under spin-$\frac52$
diffeomorphisms is given by the Einstein tensor
\begin{align}
  G_{ij} &= \varepsilon_{ikm} \varepsilon_{jln} \partial^k \partial^l \psi^{mn}
  \\
         &= \delta_{ij} \left(\Delta \bar{\psi}  -  \partial^k \partial^l \psi_{kl}\right)
           +  2  \partial_{(i} \partial^k \psi_{j)k}
           -  \Delta \psi_{ij}  - \partial_i \partial_j \bar{\psi} 
\end{align}
and its derivatives.
The variation of this tensor under a Weyl transformation is given by
\begin{align}
  \Gamma G_{ij}
  &=2 \varepsilon_{(i|km} \partial^k \gamma^{m} \mu_{|j)}
\end{align}
where $\mu_{j}=\varepsilon_{jln}  \partial^l \lambda^{n}$ is the Einstein tensor of $\lambda^{n}$.
For its traces, this implies
\begin{align}
  \Gamma \slashed{G}_{i} &= -3 \gamma_{5}\gamma_{0}\slashed \pd \mu_{i} + \varepsilon_{ijk} \pd^{k} \mu^{m} \, \\
\Gamma \bar G  &= \Gamma \delta^{ij} G_{ij} = 2 \varepsilon_{ijk} \partial^{i}\gamma^{j}\mu^{k} \,.
\end{align}
The Schouten tensor is a combination of the Einstein tensor and its
traces, i.e., $S= G + a_{1} \bar G+ b_{0} \gamma \slashed G$. Using
this ansatz and the condition that the Schouten tensor should vary to
a symmetrized derivative leads to\footnote{Note that the Schouten
  tensor used in \cite{Bunster:2014fca} is the Schouten tensor for a
  spin-$2$ field with spinor indices treated as spectator indices. The
  definition adopted here, which is different, is more adapted to the
  spin-$\frac52$ case. The tensor (\ref{eq:DefOfS}) enjoys indeed more
  useful properties.}
\begin{align} \label{eq:DefOfS}
S_{ij} &= G_{ij}  -  \frac{1}{4}  \delta_{ij} \bar G - \frac{1}{2}  \gamma_{(i} \slashed{G}_{j)}   \, . 
\end{align}
Indeed, it varies to 
\begin{equation}
\Gamma S_{ij} = \partial_{(i} \nu_{j)}
\end{equation}
where we have defined
\begin{align}
  \label{eq:nuofmu}
  \nu_i &= - \frac{1}{2} \varepsilon_{ijk} \gamma^j \mu^k + \frac{3}{2} \gamma_5 \gamma_0 \mu_i \,.
\end{align}
As in the spin-$\frac32$ case, the relation between the Einstein and
Schouten tensors is invertible and we have
\begin{equation} \label{eq:GofS}
  G_{ij} = S_{ij}  -  2  \gamma_{(i} \slashed{S}_{j)}  -  \delta_{ij} \bar S \, ,
\end{equation}
which implies $\slashed{G}_i = - 4 \slashed{S}_i$ and
$\bar G = -4 \bar S$ for the traces. Since the Einstein tensor is
identically divergenceless, the Schouten tensor satisfies
$0 = \partial^j S_{ij} - \slashed{\partial} \slashed{S}_{i} -\gamma_i
\partial_j \slashed{S}^j - \partial_{i}\bar S$. Taking the gamma-trace
of this expression gives $\partial_i \slashed{S}^i = 0$, which then
implies the Bianchi identity for the Schouten tensor
\begin{equation}
  \label{eq:Us52}
U_i[S] \equiv \partial^j S_{ij} - \slashed{\partial} \slashed{S}_{i} - \partial_{i}\bar  S = 0 \, ,
\end{equation}
which is equivalent to the divergencelessness of the Einstein tensor.
Likewise, the relation between $\mu_i$ and $\nu_i$ can be inverted to
\begin{equation} \label{eq:muofnu}
\mu^i = - \frac{1}{2} \varepsilon^{ijk} \gamma_j \nu_k - \gamma_5 \gamma_0 \nu_i \, ,
\end{equation}
and the property $\pd_i \mu^i = 0$ is equivalent to the identity
$\pd_i \nu^i - \frac{1}{2} \gamma_5 \gamma_0 \varepsilon^{ijk} \pd^i
\gamma_j \nu_k = 0$ satisfied by $\nu_i$, which can also be rewritten
as
\begin{equation}
I[\nu] \equiv \pd_i \nu^i + \slashed{\partial} \slashed{\nu} = 0 \, .
\end{equation}
An important property is that
\begin{equation}
U_i[\pd\nu] = - \frac{1}{2}\pd_i I[\nu] \, ,
\end{equation}
which shows that the Bianchi identity for the Schouten tensor is
compatible with Weyl transformations.

The conformal invariant is then the Einstein tensor of the Schouten
tensor, named the ``Cotton tensor''
\begin{align}
  D_{ij} &= \varepsilon_{ikm} \varepsilon_{jln} \partial^k \partial^l S^{mn} \\
  		 &=  2 \partial_{(i} \partial^k S_{j)k} - \Delta S_{ij} - \partial_i \partial_j \bar S \, .
\end{align}
It is again divergenceless and gamma-traceless, and invariant under higher-spin diffeomorphisms and Weyl transformations. Its explicit form in terms of the fourth derivatives of $\psi_{ij}$ is
\begin{align}
  D_{ij} &=  \Delta^2 \left(\psi_{ij}   - \frac{1}{2}  \gamma_{(i} \slashed{\psi}_{j)} - \frac{1}{4} \delta_{ij} \bar{\psi} \right) \nonumber \\
         &\quad + \frac{\Delta}{4}\left(
           \pd_i \!\pd_j \bar{\psi} + 2 \slashed{\pd} \!\pd_{(i} \slashed{\psi}_{j)}
          +\pd^k( \delta_{ij}   \pd^l \psi_{lk} - 10  \pd_{(i} \psi_{j)k}
            +2 \gamma_{(i}  \slashed{\pd} \psi_{j)k} + 2 \pd_{(i} \gamma_{j)}   \slashed{\psi}_k )
           \right) \nonumber \\
         &\quad  + \frac{1}{4} \pd_i \!\pd_j \left( 5 \pd^k \!\pd^l \psi_{kl} - 2 \slashed{\pd} \!\pd^k \slashed{\psi}_k \right)
           - \frac{1}{2} \pd_{(i} \gamma_{j)}  \pd^k \!\pd^l \slashed{\pd} \psi_{kl} \, ,
\end{align}
an expression that would have been of course very difficult to guess
(and to generalize) without the systematic construction using Einstein
and Schouten tensors.

We now turn to the proof of the two important theorems concerning the
Cotton tensor.
\paragraph{Gauge completeness.}
\label{sec:gauge52}

We first want to show that the Cotton tensor fully characterizes the
spin-$\frac{5}{2}$ diffeomorphism and Weyl-invariance. For that, we
need to show that the condition $D_{ij} = 0$ is equivalent to
$\psi_{ij}$ being pure gauge.

If $\psi_{ij}$ is pure gauge, then the Cotton vanishes by
construction, $D_{ij} = 0$. Conversely, if $D_{ij} = 0$, the Schouten
satisfies $\partial_{[i} S_{j][k,l]}=0$ or, in index-free notation,
$d_{(2)}^2 S = 0$. Using the Poincaré lemma for two-column Young
tableaux~\cite{DuboisViolette:1999rd,DuboisViolette:2001jk}, this
implies that $S = d_{(2)} \nu$ for some vector $\nu$, i.e.,
$S_{ij} = \partial_{(i} \nu_{j)}$.

Defining $G^{ij}$ and $\mu^i$ through $S_{ij}$ and $\nu_i$ by
equations \eqref{eq:GofS} and \eqref{eq:muofnu} gives
$G_{ij} = 2 \varepsilon_{(i|ab} \partial^a \gamma^b \mu_{|j)}$. It is
proven below that the ambiguity in $\nu_i$ allows us to fix
$\partial_i \mu^i = 0$. This implies
$\mu^i = \varepsilon^{ijk} \partial_j \lambda_k$ for some $\lambda_k$
and
$G_{ij} = \varepsilon_{ikm} \varepsilon_{jln} \partial^k \partial^l (
2 \gamma^{(m} \lambda^{n)} )$, or
$\varepsilon_{ikm} \varepsilon_{jln} \partial^k \partial^l ( \psi^{mn}
- 2 \gamma^{(m} \lambda^{n)} )=0$. Again using the relevant Poincaré
lemma, this implies
$\psi^{mn} = 2 \partial^{(m} \xi^{n)} + 2 \gamma^{(m} \lambda^{n)}$
for some $\xi^n$, which shows that $\psi^{mn}$ is pure gauge.

The only extra step with respect to the spin $\frac{3}{2}$-case
consists thus in establishing that the ambiguity in $\nu_i$ allows us
to fix $\pd_{i}\mu^{i} = 0$ or, equivalently, $I[\nu] = 0$. Due to the
Bianchi identity $U[S] = 0$ satisfied by the Schouten tensor, we know
that $I[\nu]$ satisfies $\pd_{i} I[\nu] =0$, which leads to
$I[\nu] = p^{(0)}$ with constant $p^{(0)}$. We can also redefine
$\nu_{i}$ as $\nu_i \sim \nu_{i}+\tilde \nu_{i}$ without changing the
Schouten tensor, as long as $\tilde{\nu}_i$ satisfies
$\pd_{(i}\tilde\nu_{j)}=0$. This is just the Killing equation for flat
space (with a spectator spinor index) and is solved by
$ \tilde \nu_{i} = q_{i}^{(0)} + q_{ij}^{(1)} x^{j}$, where
$q_{i}^{(0)}$ is a constant vector-spinor and $q_{ij}^{(1)}$ is
antisymmetric. A short computation then shows that
\begin{align}
  I[\nu+\tilde\nu]= p^{(0)} - \gamma^{ij}q^{(1)}_{ij} \, .
\end{align}
Therefore, choosing $q^{(1)}_{ij} = - \frac{1}{6}\gamma_{ij}p$ fixes
$I[\nu + \tilde{\nu}] = 0$, thus concluding the proof.

\paragraph{Conformal Poincaré lemma.}

The Cotton tensor is symmetric, divergenceless and gamma-traceless. We
will now prove that any tensor with these properties, i.e., any
symmetric tensor $T_{ij}$ satisfying $\pd^{i} T_{ij} = 0$ and
$\slashed{T}_{i} = 0$, can be written as the Cotton tensor
$T_{ij} = D_{ij}[\psi]$ for some $\psi_{ij}$.

Divergenceless of a symmetric tensor implies, using the generalized
Poincaré lemma~\cite{DuboisViolette:1999rd,DuboisViolette:2001jk},
that $T_{ij}$ is the Einstein tensor of some symmetric tensor
$S_{ij}$,
\begin{equation}
T_{ij}= \varepsilon_{ikm} \varepsilon_{jln} \partial^k \partial^l S^{mn}   \,.
\end{equation}
The condition $\slashed T_{i}=0$ leads to
\begin{align}
0 &= \varepsilon_{ijk}\pd^{j} (\pd_{l} S^{lk} - \slashed \pd \slashed S^{k}-\pd^{k}\bar S) \\
&= \varepsilon_{ijk}\pd^{j} U^k[S] \, ,
\end{align}
where we have added the last trace-term for convenience, at no cost
since partial derivatives commute while
$\varepsilon_{ijk} = - \varepsilon_{ikj}$.
The Poincaré lemma then implies
\begin{align}
  \label{eq:coneq}
U_i[S] = \partial^j S_{ij} - \slashed{\partial} \slashed{S}_{i} - \partial_{i}\bar  S = \pd_{i}\rho \,.
\end{align}
Suppose for now that the right hand side vanishes. Then, our tensor
$S$ satisfies the Bianchi identity \eqref{eq:Us52} for the Schouten
tensor of that spin. Therefore, the tensor $G_{ij}$ defined by
\begin{equation}
  G_{ij} = S_{ij}  -  2  \gamma_{(i} \slashed{S}_{j)}  -  \delta_{ij} \bar S \, ,
\end{equation}
satisfies $\pd^{i}G_{ij}=0$, as is proven in the beginning of this
section. This followed from the invertibility of the definition of the
Schouten in terms of the Einstein tensor, for which the definition in
terms of $\psi$ was irrelevant. Now, since $G$ is divergenceless, we
have
\begin{equation}
  G_{ij} = \varepsilon_{ikm} \varepsilon_{jln} \partial^k \partial^l \psi^{mn}
\end{equation}
for some symmetric $\psi^{mn}$, which shows that $S$ is the Schouten
tensor of $\psi$ and therefore that $T_{ij}$ is its Cotton, $T_{ij} = D_{ij}[\psi]$.

To finish the proof we need to show that we can use the ambiguities in
the Poincaré lemma to indeed set the right hand side of
\eqref{eq:coneq} to zero. The freedom we have is given by
$S_{ij} \sim S_{ij} + \pd_{(i}\nu_{j)}$ which leads to
\begin{align}
  U_{i}[\pd \nu]=-\frac{1}{2}\pd_{i}(\pd_{j} \nu^{j} + \slashed \pd\slashed \nu)=\pd_{i}\rho \,.
\end{align}
Therefore, it is sufficient to choose $\nu$ such that
\begin{align}\label{eq:Inurho}
I[\nu]= \pd_{i} \nu^{i} + \slashed \pd\slashed \nu=-2\rho \,.
\end{align}
To prove that such a $\nu_i$ exists, we define a vector-spinor $\mu^i$
in terms of $\nu^i$ as in equation \eqref{eq:muofnu} and use the
invertibility of that relation. Equation \eqref{eq:Inurho} is then
equivalent to
\begin{align}
\pd_{i}\mu^{i} = \hat \rho \, ,
\end{align}
where $\hat{\rho} = \gamma_5 \gamma_0 \rho$, which always has a
solution. This therefore completes the proof.

\subsection{Schouten and Cotton tensors: general half-integer spin}
\label{sec:general-half-integer}

The spin-$(s + \frac12)$ field is a symmetric tensor(-spinor) with $s$
vectorial indices $\psi$. Again, a complete set of invariants under
spin-$(s+ \frac12)$ diffeomorphisms is given by the Einstein tensor
\begin{equation}
G = \left(\varepsilon \cdot \partial \, \cdot \right)^s  \psi \, .
\end{equation}
The variation of this tensor under a Weyl transformation
\eqref{Transf1a} is
\begin{equation}
\Gamma G = s  \left(\varepsilon \cdot \partial \cdot \gamma \right) \mu \, ,
\end{equation}
where $\mu=(\epsilon \cdot \partial \, \cdot)^{s-1}\lambda$ is the
Einstein tensor of $\lambda$ or, equivalently, any symmetric
divergenceless tensor of rank $s-1$. This implies, for the $p$-th
trace and gamma-trace of $G$
\begin{align}
  \Gamma G^{[p]}
  &=
    2p  (\varepsilon \cdot \partial \cdot \gamma \, \cdot) \mu^{[p-1]} +  \left(s - 2p \right) \left(\varepsilon \cdot \partial \cdot \gamma \right) \mu^{[p]} &  (0 &\leq p \leq \lfloor s/2 \rfloor)\, ,
  \\
  \Gamma \slashed{G}^{[p]}
  &=
    - \gamma_5 \gamma_0  \left(s + 1 \right) \slashed{\partial} \mu^{[p]} + \left(s - 1 - 2p\right) \left(\varepsilon \cdot \partial \, \cdot \right) \mu^{[p]} & (0&\leq p \leq \lfloor (s-1)/2 \rfloor)\, ,
\end{align}
where $\lfloor s/2 \rfloor$ denotes the largest integer equal or
smaller than $s/2$. The Schouten tensor will be built out of these
quantities multiplied by $p$ delta functions and an additional gamma
matrix for $\slashed G^{[p]}$. The variation of these terms is given
by
\begin{align}
  \delta^p \Gamma G^{[p]} &=
                            2p \delta^p (\varepsilon \cdot \partial \cdot \gamma \, \cdot) \mu^{[p-1]}
                            + \left(s - 2p \right) \delta^p \left(\varepsilon \cdot \partial \cdot \gamma  \right) \mu^{[p]} \, ,
  \\
  \delta^p \gamma \Gamma \slashed{G}^{[p]} &= 
                                             - \gamma_5 \gamma_0 \left(s + 1 \right) \delta^p \partial \mu^{[p]} + \left(s + 1 \right) \delta^p \left(\varepsilon \cdot \partial \cdot \gamma  \right) \mu^{[p]} \nonumber
  \\
                                           &\quad + \left(s - 1 - 2p\right) \delta^p \gamma \left(\varepsilon \cdot \partial \,\cdot \right) \mu^{[p]} \, .
\end{align}
At first sight, it does not seem possible to combine these expressions
in order to obtain a symmetrized derivative, but we have to take into
account the following identity
\begin{align}
0 &= 4  \varepsilon_{[ijk} \delta_{l]m} \partial^i \gamma^j \mu^k
  \\
  &=
\varepsilon_{ijk} \delta_{lm} \partial^i \gamma^j \mu^k
     -  \varepsilon_{jkl}  \partial_m \gamma^j \mu^k
     +  \varepsilon_{kli} \partial^i \gamma_m \mu^k
     -  \varepsilon_{lij} \partial^i \gamma^j \mu^m  \, ,
\end{align}
where the spectator indices of $\mu$ have been left unwritten. After
symmetrization in $lm$ (together with the remaining indices of $\mu$),
this gives
\begin{equation}
  0
  =
  \delta \varepsilon \cdot \partial \cdot \gamma \cdot \mu
   -  \partial \left( \varepsilon \cdot \gamma \,\cdot \right) \mu
   +  \gamma\left(\varepsilon \cdot \partial \, \cdot \right) \mu
   -  \left(\varepsilon \cdot \partial \cdot \gamma  \right) \mu \, ,
\end{equation}
which in turn implies
\begin{align}
  \delta^p \Gamma G^{[p]}
  &= 
    2p  \delta^{p-1} \partial \left(\varepsilon \cdot \gamma \,\cdot \right) \mu^{[p-1]}
     -  2p  \delta^{p-1}\gamma\left(\varepsilon \cdot \partial \, \cdot \right) \mu^{[p-1]}
     +  2p  \delta^{p-1}\left(\varepsilon \cdot \partial \cdot \gamma \right) \mu^{[p-1]}
    \nonumber \\
  &\quad
    +  \left(s - 2p \right) \delta^p \left(\varepsilon \cdot \partial \cdot \gamma  \right) \mu^{[p]} \, .
\end{align}
This leads us to define the Schouten tensor as
\begin{equation}
  \label{eq:schoutendef}
  S =
  \sum_{p = 0}^{\lfloor s/2 \rfloor} a_p \delta^p G^{[p]} 
  + \sum_{p = 0}^{\lfloor (s-1)/2 \rfloor} b_p \delta^p \gamma \slashed{G}^{[p]} \, .
\end{equation}
 Requiring the gauge variation of the Schouten to
be a symmetrized derivative ($\Gamma S = \partial \nu$ for some
symmetric tensor $\nu$) imposes
\begin{align}
0 &= -  2 \left(p + 1\right) a_{p+1}  +  \left(s - 1 - 2p\right) b_p \, ,
\\
0 &= 2 \left(p + 1 \right) a_{p+1} +  \left(s - 2p\right) a_p  +  \left(s + 1\right) b_p \, .
\end{align}
Taking the initial condition $a_0 = 1$, the solution to these
recurrence relations is
\begin{align}
  a_p &= \frac{(-1)^p}{4^p\,p!}\frac{\left(s-p\right)!}{\left(s-2p\right)!} \, , \label{eq:ap}
  \\
  b_p & =- \frac{1}{2} \frac{(-1)^p}{4^p\,p!} \frac{\left(s-p-1\right)!}{\left(s-2p-1\right)!}=- \frac{1}{2} \frac{s-2p}{s-p} a_{p}\label{eq:bp} \, .
\end{align}
The gauge variation of the Schouten tensor is then indeed a
gradient and reads explicitly
\begin{equation}
  \Gamma S = \partial \nu
\end{equation}
for a symmetric tensor $\nu$ which is related to $\mu$ as
\begin{equation}
  \nu = \sum_{p = 0}^{\lfloor s/2 \rfloor} 2 p \, a_p \, \delta^{p-1} (\varepsilon \cdot \gamma \, \cdot ) \mu^{[p-1]} - \sum_{p = 0}^{\lfloor (s-1)/2 \rfloor} b_p\, (s+1) \gamma_5 \gamma_0 \,\delta^p \mu^{[p]} \, . 
\end{equation}
The Schouten tensor satisfies the Bianchi identity
\begin{equation}\label{eq:UofS}
  U[S]\equiv \pd \cdot S - \slashed{\pd} \slashed{S} - (s-1) \pd \bar{S} = 0 \, ,
\end{equation}
which is equivalent to the divergencelessness $\pd \cdot G = 0$ of the
Einstein tensor. Indeed, plugging formula \eqref{eq:schoutendef} into
this identity and using the form of the $a_p$, $b_p$ coefficients, one
gets
\begin{equation}
  U[S] = \frac{1}{s} \left[
    \sum_{p = 0}^{\lfloor s/2 \rfloor} a_p (s-2p) \delta^p \pd \cdot G^{[p]} 
    +\sum_{p = 0}^{\lfloor (s-1)/2 \rfloor} b_p (s-2p-1) \delta^p \gamma \pd \cdot \slashed{G}^{[p]} \right] \, ,
\end{equation}
which vanishes by virtue of $\pd \cdot G = 0$. Similarly, on the
parameter $\nu$ for the gauge transformations of the Schouten tensor,
the identity equivalent to $\pd \cdot \mu = 0$ is
\begin{equation}
  \label{eq:Iofnu}
  I[\nu] \equiv \pd \cdot \nu + \slashed{\pd} \slashed{\nu} + (s-2) \pd \bar{\nu} = 0
\end{equation}
or, equivalently (using
$\gamma^i \gamma^j = \delta^{ij} + \varepsilon^{ijk} \gamma_k \gamma_5
\gamma_0$),
\begin{equation}
  \pd \cdot \nu - \frac{1}{2} \gamma_5 \gamma_0 (\varepsilon \cdot \pd \cdot \gamma \cdot \nu) + \frac{(s-2)}{2} \pd \bar{\nu} = 0 \, .
\end{equation}
These identities are compatible since
\begin{equation}
  \label{eq:Udnu}
  U[\pd \nu] = - \frac{(s-1)}{s} \pd I[\nu] \, ,
\end{equation}
which guarantees (as it should!) that the property $U[S] = 0$ is not
destroyed by a Weyl transformation.

The Einstein tensor of the Schouten tensor is then the
searched-for conformally invariant Cotton tensor
\begin{equation}
  D = \left(\varepsilon \cdot \partial \, \cdot \right)^s S \, .
\end{equation}
It is obviously symmetric, divergenceless and invariant under the full
gauge and Weyl transformations of the field. It is also
gamma-traceless owing to the identity $U[S] = 0$; indeed, a short
computation shows that
\begin{equation}
  \slashed{D} = (\varepsilon \cdot \pd \, \cdot)^{s-1} U[S]=0\, .
\end{equation}

As in the spin-$\frac{5}{2}$ case, the proofs of the gauge
completeness and the conformal Poincaré lemma heavily rely on the
identities \eqref{eq:UofS} and \eqref{eq:Iofnu} satisfied by the
Schouten tensor and the $\nu$ parameter.

\subsubsection{Gauge completeness}
\label{sec:gauge-completeness}

If $\psi$ is pure gauge, then $D = 0$ by construction. To prove the
converse, we proceed as before, using the differential $d_{(s)}$. If
$D = 0$, the Schouten tensor satisfies $d_{(s)}^s S = 0$, which
implies $S = d_{(s)} \nu$ for some rank $s-1$ symmetric tensor
$\nu$. Defining $G$ and $\mu$ by inverting the definitions above, this
is equivalent to
$G = s (\varepsilon \cdot \partial \cdot \gamma \, \cdot) \mu$. Now, if the
ambiguity in $\nu$ allows us to fix $\partial \cdot \mu = 0$, implying
that $\mu$ is the Einstein tensor of some $\lambda$, we get
$G[\psi] = s ( \varepsilon \cdot \partial \, \cdot)^s (\gamma \lambda)$, or
$( \varepsilon \cdot \partial \, \cdot)^s (\psi - s \, \gamma \lambda) = 0$.
Again using the Poincaré lemma for $d_{(s)}$, this implies that
$\psi = s ( \partial \xi + \gamma \lambda )$ for some $\xi$, showing
that $\psi$ is pure gauge.

The key step above is thus again that the ambiguity in $\nu$ allows us
to fix $\pd\cdot\mu = 0$ or, equivalently, $I[\nu] = 0$ (as defined in
\eqref{eq:Iofnu}). This can be seen as follows. First of all, the
Bianchi identity for the Schouten tensor implies that $I[\nu]$
satisfies $\pd I[\nu] = 0$ or, in index notation,
\begin{equation}
\pd_{(i_1} I_{i_2 \cdots i_{s-1})}[\nu] = 0\, .
\end{equation}
The general solution of this equation is \cite{Thompson86,Wolf98,McLenaghan04}
\begin{equation}
  I_{i_1 \cdots i_{s-2}} = \sum_{n=0}^{s-2} p^{(n)}_{i_1 \cdots i_{s-2} \, j_1 \cdots j_n} x^{j_1} \cdots x^{j_n}\, ,
\end{equation}
where the $p^{(n)}$ are constant tensor with $(s-2, n)$ Young
symmetry (in the symmetric convention), i.e.,
\begin{align}
  p^{(n)}_{i_1 \cdots i_{s-2} \, j_1 \cdots j_n} &= p^{(n)}_{(i_1 \cdots i_{s-2}) \, j_1 \cdots j_n} = p^{(n)}_{i_1 \cdots i_{s-2} \, (j_1 \cdots j_n)} \, ,
  \\
  p^{(n)}_{(i_1 \cdots i_{s-2} \, j_1) j_2\cdots j_n} &= 0 \, .
\end{align}

On the other hand, the ambiguity in $\nu$ is given by solutions of
$\pd \tilde{\nu} = 0$, i.e.,
\begin{equation}
  \tilde{\nu}_{i_1 \cdots i_{s-1}} = \sum_{n=0}^{s-1} q^{(n)}_{i_1 \cdots i_{s-1} \, j_1 \cdots j_n} x^{j_1} \cdots x^{j_n} \, ,
\end{equation}
where the $q^{(n)}$ are constant tensors with $(s-1, n)$ Young
symmetry (again in the symmetric convention). Now, computing
$I[\tilde{\nu}]$ shows that we can use this ambiguity to fix
$I[\nu] = 0$ provided we choose the tensors $q^{(n)}$ such that
\begin{equation}
\gamma^{rs} q^{(n+1)}_{i_1 \cdots i_{s-2} r \, j_1 \cdots j_n s} = p^{(n)}_{i_1 \cdots i_{s-2} \, j_1 \cdots j_n}
\label{eq:EqForQ}
\end{equation}
(up to factors that can be absorbed in the $q^{(n)}$). That this
equation always possesses a solution for arbitrarily given $p$'s with
the $(s-2, n)$ Young symmetry is proven in Appendix
\ref{sec:proof-gamma-q}.

\subsubsection{Conformal Poincaré lemma}
\label{sec:conf-poinc-lemma}

As was mentioned above, the Cotton tensor is symmetric, divergenceless
and gamma-traceless. We now want to prove the converse, i.e., that for
any symmetric tensor $T$ satisfying $\pd \cdot T = 0$ and
$\slashed{T} = 0$, there exists some $\psi$ such that $T = D[\psi]$.
First of all, $\pd \cdot T = 0$ implies that there exists a symmetric
tensor $S$ of which $T$ is the Einstein tensor,
\begin{equation}
T = G[S]\, .
\end{equation}
Then, due to $\slashed{T} = 0$, this tensor satisfies
$(\varepsilon \cdot \pd \, \cdot)^{s-1} U[S] = 0$, which implies
\begin{equation}
U[S] = \pd \rho
\end{equation}
for some symmetric tensor $\rho$ using the appropriate Poincaré
lemma. Now, we would like to use the ambiguity in $S$ to cancel $\rho$
so that $U[S] = 0$. Indeed, this would imply that $S$ satisfies the
Bianchi identity of the Schouten tensor and therefore that some $\psi$
exists such that $S = S[\psi]$, which shows that $T = D[\psi]$.

The ambiguity in $S$ is given by $S \sim S + \pd \nu$ (since it is
only defined through its Einstein tensor). Because of equation
\eqref{eq:Udnu}, we can fix $U[S] = 0$ provided we can solve the
differential equation
\begin{equation}
I[\nu] = \rho
\end{equation}
(up to factors and signs that can be absorbed in $\nu$). Because of
the invertible relation between $\nu$ and $\mu$ (or between $I[\nu]$
and $\pd \cdot \mu$), this is equivalent to
\begin{equation}
\pd\cdot \mu = \hat\rho\, ,
\end{equation}
where $\hat\rho$ is an invertible combination of $\rho$ and its traces
and gamma-traces. This equation can be solved for $\mu$, which
finishes the proof.

\section{Equations of motion as twisted self-duality}
\label{Hamiltonian}

In this section, we rewrite the equations of motion of the fermionic
spin $(s + \frac{1}{2})$-field in four spacetime dimensions as twisted
self-duality conditions on the curvature, supplemented by a purely
spatial constraint.

\subsection{From the Fronsdal to the Riemann tensor}
\label{sec:from-fronsd-riem}

The equations of motion for the spin $(s+\frac{1}{2})$-field
$\psi_{\mu_1 \cdots \mu_s}$ are of first-order in derivatives and
given by
\begin{equation}
\cF_{\mu_1 \cdots \mu_s} = 0\, ,
\end{equation}
where the Fronsdal tensor is defined as
\begin{equation} \label{eq:fronsdaltensor}
\cF_{\mu_1 \cdots \mu_s} = \bslashed{\pd} \psi_{\mu_1 \cdots \mu_s} - s \pd_{(\mu_1} \bslashed{\psi}_{\,\mu_2 \cdots \mu_s)}
\end{equation}
and the field itself satisfies the trace condition
$\bslashed{\psi}^{\ \prime} = 0$ (see Section \ref{sec:introduction}
for the notation). Under a gauge transformation
\begin{equation}\label{eq:psigauge}
\Gamma \psi_{\mu_1 \cdots \mu_s} = s \pd_{(\mu_1} \xi_{\mu_2 \cdots \mu_s)}\, ,
\end{equation}
the Fronsdal tensor transforms as
\begin{equation}
\Gamma \cF_{\mu_1 \cdots \mu_s} = -s(s-1) \pd_{(\mu_1} \pd_{\mu_2} \bslashed{\xi}_{\mu_3 \cdots \mu_s)}
\end{equation}
or, in index-free notation, $\Gamma \cF = d_{(s)}^2 \bslashed{\xi}$,
where the differential $d_{(s)}$ satisfies $d^{s+1}_{(s)} = 0$. The
equations of motion are gauge-invariant if the gauge parameter is
gamma-traceless.

The first step is to rewrite these equations in a manner that is
invariant under the larger set of traceful gauge transformations; this
is done along the lines of \cite{Damour:1987vm,Bekaert:2003az} by
going to the (spacetime) Riemann tensor. This formulation contains
more derivatives of the field, and the original formulation can then
be recovered by fixing the gauge. The Riemann tensor is the
$(s,s)$-tensor
\begin{equation}
R_{\mu_1\nu_1 \cdots \mu_s \nu_s} = 2^s \pd_{[\nu_1|} \cdots \pd_{[\nu_s|} \psi_{|\mu_1] \cdots |\mu_s]} \,.
\end{equation}
It is invariant under \eqref{eq:psigauge} even if the trace parameter
is traceful. Taking the gamma-trace of $R$, we get the $(s,s-1)$-tensor
\begin{align}
  \gamma^{\nu_1} R_{\mu_1\nu_1 \cdots \mu_s \nu_s}
  &= 2^s \pd_{\nu_2} \cdots \pd_{\nu_{s}} \left( \bslashed{\pd} \psi_{\mu_1 \mu_2 \cdots \mu_s} - \pd_{\mu_1} \bslashed{\psi}_{\,\mu_2 \cdots \mu_s} \right) \\
  &= 2^s \pd_{\nu_2} \cdots \pd_{\nu_{s}} \left( \bslashed{\pd} \psi_{\mu_1 \mu_2 \cdots \mu_s} - s \pd_{(\mu_1} \bslashed{\psi}_{\,\mu_2 \cdots \mu_s)} \right) \\
  &= 2^s \pd_{\nu_2} \cdots \pd_{\nu_{s}} \cF_{\mu_1 \mu_2 \cdots \mu_s}
\end{align}
where the obvious antisymmetrization in $\mu_k, \nu_k$ ($k\geq 2$) are
not written explicitly to avoid cluttered notation. We have added the
necessary terms in the second line (at no cost since partial
derivatives commute) to make the Fronsdal tensor appear. In index-free
notation, this is
\begin{equation}
  \bslashed{R} = d_{(s)}^{s-1} \cF \, .
\end{equation}
Another way to understand why a relation of this type must exist is
that, because of the gauge transformation property of the Fronsdal
tensor, the quantity $d_{(s)}^{s-1} \cF$ is gauge-invariant and must
therefore, like any other local gauge invariant function,
be expressible in terms of the Riemann tensor.

Now, $\cF = 0$ implies $\bslashed{R} = 0$; conversely,
$\bslashed{R} = 0$ implies $\cF = d^2_{(s)} \zeta$ using the relevant
Poincaré lemma, for some $\zeta$ that we can always write as
$\zeta = \bslashed{\xi}$. This is the equation $\cF = 0$ up to a
traceful gauge transformation. Thus, one can reach $\cF = 0$ by a
gauge transformation when $\bslashed{R} = 0$. One can further show
\cite{Francia:2002pt} that there is enough gauge freedom (when
$\bslashed{R} = 0$) to impose also the triple gamma-trace condition on
the field $\psi$ itself.

\subsection{Rewriting as a twisted self-duality condition}
\label{sec:rewriting-as-twisted}

We show in this subsection that the geometrical equation
$\bslashed{R} = 0$ is equivalent to the system
\begin{align}\label{eq:tsd}
R &= - \gamma_5 \, \hs R \, , & \gamma^{kl} R_{kl\, i_2 j_2\, \cdots\, i_s j_s} &= 0 \, ,
\end{align}
i.e., the twisted self-duality condition supplemented by a constraint
on the purely spatial components.\footnote{In our conventions
  $(\gamma_5)^2 = -1$ which is consistent with $(\hs)^2 = -1$.} This
is the analog for fermionic fields of the twisted self-duality
condition derived in \cite{Henneaux:2016zlu} for bosonic fields.

\subsubsection*{Spin-$\frac{3}{2}$}

We first start with the spin-$\frac{3}{2}$ case, which
  illustrates the main points. The equation $\bslashed{R} = 0$ is in
this case
\begin{align}
\gamma^\mu R_{\mu\nu} &= 0 \, , & R_{\mu\nu} &= \pd_{\nu} \psi_\mu - \pd_\mu \psi_\nu \, .
\end{align}
It is equivalent to the the usual Rarita-Schwinger equation $\gamma^{\mu\nu\rho} R_{\nu\rho} = 0$.

\begin{itemize}
\item $\bslashed{R}=0\, \Rightarrow \, R = - \gamma_5 \, \hs R \, \text{ and } \, \gamma^{kl} R_{kl\, i_2 j_2\, \cdots\, i_s j_s} = 0$:

  First, by contracting with $\gamma^\nu$, one sees that
  $\gamma^\mu R_{\mu\nu} = 0$ implies $\gamma^{\mu\nu} R_{\mu\nu} = 0$
  (since $R_{\mu\nu}$ is antisymmetric). Splitting time and
  space, this is $2\gamma^0\gamma^i R_{0i} + \gamma^{ij} R_{ij} = 0$.
  The first term is the $0$ component of $\bslashed{R}=0$ and
  therefore vanishes; this shows that $\bslashed{R}=0$ indeed implies
  the spatial constraint $\gamma^{ij} R_{ij} = 0$.

  Then, using the gamma matrix identities
  \begin{align}
    \gamma_{\mu\nu} \gamma^{\rho\sigma} &= \gamma\indices{_{\mu\nu}^{\rho\sigma}} - 4 \delta^{[\rho}_{[\mu} \gamma\indices{_{\nu]}^{\sigma]}} - 2 \delta_{\mu\nu}^{\rho\sigma} \\
    \gamma^{\mu\nu\rho} \gamma^\sigma &= \gamma^{\mu\nu\rho\sigma} + 3 \gamma^{[\mu\nu} \eta^{\rho]\sigma}
  \end{align}
(valid in all dimensions) and
\begin{equation}
\gamma_{\mu\nu\rho\sigma} = \varepsilon_{\mu\nu\rho\sigma} \gamma_5
\end{equation}
with $\varepsilon_{0123} = +1$ (specific to four dimensions), we get
\begin{align}
0 &= \gamma\indices{_{\mu\nu}^\rho} \gamma^{\sigma} R_{\rho\sigma} = \varepsilon_{\mu\nu\rho\sigma} \gamma_5 R^{\rho\sigma} + 2 \gamma\indices{_{[\mu}^\sigma} R_{\nu]\sigma} \\
0 &= \gamma_{\mu\nu} \gamma^{\rho\sigma} R_{\rho\sigma} = \varepsilon_{\mu\nu\rho\sigma} \gamma_5 R^{\rho\sigma} + 4 \gamma\indices{_{[\mu}^\sigma} R_{\nu]\sigma} - 2 R_{\mu\nu} \, .
\end{align}
Taken together, these two equations imply
\begin{equation}
R_{\mu\nu} = - \frac{1}{2} \varepsilon_{\mu\nu\rho\sigma} \gamma_5 R^{\rho\sigma}\, ,
\end{equation}
which is the twisted self-duality in components.

\item $ R = - \gamma_5 \, \hs R \, \text{ and } \, \gamma^{kl} R_{kl\, i_2 j_2\, \cdots\, i_s j_s} = 0 \, \Rightarrow \, \bslashed{R}=0$:

Splitting space and time, the twisted self-duality is
\begin{equation}
R_{0i} = -\frac{1}{2} \varepsilon_{ijk} \gamma_5 R^{jk} \, .
\end{equation}
Contracting with $\gamma^i$ and using the identity $\gamma_{jk} = -\varepsilon_{ijk} \gamma^i \gamma_0 \gamma_5$, this gives
\begin{equation}
2\gamma^0\gamma^i R_{0i} - \gamma^{ij} R_{ij} = 0\, .
\end{equation}
Using the constraint, this reduces to the zero component of
$\bslashed{R} = 0$. We are still missing the spatial components of
that equation, which read $\gamma^0 R_{0i} + \gamma^k R_{ki} = 0$.
This is proved by using the identity
\begin{equation}
\gamma_i \gamma_{jk} = \gamma_{ijk} + 2 \delta_{i[j} \gamma_{k]} = - \varepsilon_{ijk} \gamma_0 \gamma_5 + 2 \delta_{i[j} \gamma_{k]}\, ,
\end{equation}
which gives, using the constraint and the twisted self-duality condition
\begin{align}
0 &= \gamma_i \gamma_{jk} R^{jk} = -\varepsilon_{ijk} \gamma_0 \gamma_5 R^{jk} + 2 \gamma^k R_{ik} = 2 \gamma_0 R_{0i} + 2 \gamma^k R_{ik} \\
&= - 2 \left( \gamma^0 R_{0i} + \gamma^k R_{ki} \right)\, .
\end{align}
\end{itemize}

\subsubsection*{Arbitrary spin}

The proof of the previous section carries over without any change if
one adds as many indices as necessary to $R$. This shows the
equivalence between $\bslashed{R} = 0$ and
\begin{align}
R &= - \gamma_5 \, \hs R \, , & \gamma^{kl} R_{kl\, \mu_2 \nu_2\, \cdots\, \mu_s \nu_s} &= 0\, ,
\end{align}
where the constraint carries additional spacetime indices. To finish
the proof in the arbitrary spin case, we therefore need to show that
the subset of these constraints with only spatial indices implies all
the others components (with one or more zeros).

Using the twisted self-duality condition on other groups of indices, we can
dualize every temporal component appearing in the constraint to
spatial indices, for example
\begin{equation}
\gamma^{kl} R_{kl\, 0i\, p_3 q_3 \cdots p_s q_s} = - \frac{1}{2} \varepsilon_{imn} \gamma^{kl} R\indices{_{kl}^{mn}_{p_3 q_3 \cdots p_s q_s}} = 0\, .
\end{equation}
This shows that the purely spatial constraint appearing in \eqref{eq:tsd} is sufficient.

\subsection{Hamiltonian constraint}

The constraint $\gamma^{kl} R_{kl\, i_2 j_2\, \cdots\, i_s j_s} = 0$
possesses an interesting interpretation in terms of the dynamics: it
is equivalent to the constraint that appears in the Hamiltonian
formulation, as we now show.

The Fang--Fronsdal action \cite{Fang:1978wz} is linear in the first
order derivatives and is thus already in Hamiltonian form (up to field
redefinitions of the variables). The components
$\psi_{0k_2k_3 \ldots k_s}$ with one index equal to $0$ are Lagrange
multipliers enforcing the ``Hamiltonian constraints'' on the dynamical
variables
\cite{Aragone:1979hw,Borde:1981gh,Bunster:2014fca,Campoleoni:2017vds}.
These constraints arise from the components with one index equal to
$0$ of  the original equations of motion $\cF = 0$,
\begin{equation}
0 =
\left( s - 1 \right) \partial_{(k_2} \Xi_{k_3\ldots k_s )} + 2 \gamma^{lm} \partial_{l} \psi_{m k_2\ldots k_s} + \left(s - 1 \right)  \partial_{(k_2} \bar{\psi}_{k_3\ldots k_s )} , \label{const}
\end{equation}
where 
\begin{align}
\label{eq:xidef}
\Xi_{k_3 \ldots k_s} &= \psi_{00k_3 \ldots k_s} -2 \gamma^{0} \gamma^{i}\psi_{0 i k_3 \ldots k_s} \, .
\end{align}
If we take $s-1$ curls of this expression (i.e., if we compute its
Einstein tensor), we will get the vanishing of the gamma-trace
of the Einstein tensor of $\psi_{k_1 \ldots k_s}$,
\begin{equation}
\slashed{G}[\psi] = 0\, .
\end{equation}
Using the spatial $\varepsilon$ tensor, this is equivalent to the
constraint written in \eqref{eq:tsd} in terms of the Riemann tensor.

\section{Prepotentials}
\label{Prepotentials}

The fermionic higher-spin conformal geometry provides the tools for
introducing prepotentials to write the action for the twisted
self-duality equations as a typical, and remarkably simple,
prepotential action. It is equivalent to the usual Fronsdal action,
where the constraints are solved.

\subsection{Solution of the constraints}

Since the Einstein tensor $G[\psi]$ is symmetric and divergenceless,
the conformal Poincaré lemma, proven in Section \ref{sec:conf-poinc-lemma},
implies that the constraint
$\slashed{G}[\psi] = 0$ is solved by writing
\begin{equation}
G[\psi] = D[\chi]  \label{eq:GandD}
\end{equation}
in terms of a prepotential $\chi$, where $D$ is the Cotton tensor. A
formula realizing this is simply
\begin{equation}
\psi[\chi] = S[\chi]\, , \label{eq:SolForChi}
\end{equation}
where $S$ is the Schouten tensor, since the Cotton tensor is exactly
defined as the Einstein of the Schouten. The simplicity of this
formula with respect to the bosonic case seems to be a recurring fact
for fermionic fields: see \cite{Henneaux:2017xsb,Lekeu:2018kul}, where
the same happens for fermionic fields in other dimensions.

Plugging this back into \eqref{const}, we get the equation
\begin{align}
0 &= (s-1)\partial \Xi +(s-1) \partial \bar{S}[\chi] - 2 \left(\partial \cdot S[\chi] - \slashed{\partial} \slashed{S}[\chi] \right)= (s-1) \partial \left( \Xi - \bar S[\chi] \right)
\end{align}
using the identity
$\partial \cdot S - \slashed{\partial} \slashed{S} -(s-1) \partial\bar
S=0$. A particular solution of this equation for $\Xi[\chi]$ is
\begin{equation}
\Xi[\chi] = \bar S[\chi] \, .
\end{equation}
In this way, all the dynamical variables (spatial components of $\psi$
and $\Xi$) are expressed in terms of the prepotential $\chi$.

We have chosen a particular solution (\ref{eq:SolForChi}) of the
equation (\ref{eq:GandD}). Given the properties of the Einstein
tensor, the most general solution will differ from
(\ref{eq:SolForChi}) by a gauge transformation of $\psi$, and so is
physically equivalent to the choice adopted here. Moreover, the
relation (\ref{eq:SolForChi}) clearly satisfies (by construction of
the Schouten tensor) the property that a Weyl transformation of $\chi$
induces a gauge transformation of $\psi$. Finally, we note that the
ambiguity in $\Xi$, for fixed $\psi$'s, is given by a solution of the
Killing tensor equation for a tensor with $s-2$ spatial
indices. Such a solution is constant or blows up at infinity, and can
be dropped if we assume that the spin-$s$ field goes to zero at
infinity.

\subsection{Equations of motion}

A sufficient subset of the twisted self-duality equations is given by the components with at most one zero, i.e.,
\begin{equation}\label{eq:spatialtsd}
R_{0s j_2 k_2 \ldots j_s k_s} = - \frac{1}{2} \varepsilon_{spq} \gamma_5 R\indices{^{pq}_{j_2 k_2 \ldots j_s k_s}}\, .
\end{equation}
This is proven indirectly below, starting from the action for the
field $\psi$. Those equations still contain temporal components of the
field; to get rid of those, we take an extra curl, giving the equation
\begin{equation}
\varepsilon^{irs} \pd_r \left( R_{0s j_2 k_2 \ldots j_s k_s} + \frac{1}{2} \varepsilon_{spq} \gamma_5 R\indices{^{pq}_{j_2 k_2 \ldots j_s k_s}} \right) = 0\, .
\end{equation}
This equation is equivalent to \eqref{eq:spatialtsd}: the missing
components can be recovered using the appropriate Poincaré lemmas. The
fact that only spatial components of $\psi$ appear in this equation
can be made more manifest by writing
$\varepsilon^{irs} \pd_r R_{0s j_2 k_2 \ldots j_s k_s} = \frac{1}{2}
\varepsilon^{irs} \dot{R}_{rs j_2 k_2 \ldots j_s k_s}$ using the
differential Bianchi identities for the Riemann tensor. Contracting
further with epsilon tensors to make the Einstein tensor of $\psi$
appear, we get
\begin{equation}
\dot{G}^{i_1 \ldots i_s}[\psi] + \gamma_5 \varepsilon^{i_1 jk} \pd_j G\indices{_k^{i_2 \ldots i_s}}[\psi] = 0\, .
\end{equation}
As emphasized above, this equation (supplemented by the
gamma-tracelessness of $G[\psi]$) is equivalent to the usual equations
of motion for $\psi$. In terms of the prepotential $\chi$, this is
\begin{equation}\label{eq:eomprepot}
\dot{D}^{i_1 \ldots i_s}[\chi] + \gamma_5 \varepsilon^{i_1 jk} \pd_j D\indices{_k^{i_2 \ldots i_s}}[\chi] = 0 \, .
\end{equation}
We have therefore succeeded in rewriting the equations of motion for
the field $\psi$ purely in terms of the prepotential $\chi$;
remarkably, they then take the form ``sum of time derivative and curl
of the Cotton tensor vanishes'' that is familiar in the prepotential
formulation
\cite{Bunster:2012km,Henneaux:2016zlu,Henneaux:2016opm,Henneaux:2017xsb,Lekeu:2018kul,Henneaux:2018rub}.

\subsection{Action in terms of prepotentials}

The equation above follows from the prepotential action
\begin{equation}\label{eq:actionprepot}
S[\chi] = - i \int \! dt \,d^3\!x\,  \chi^\dagger_{i_1 \ldots i_s} \left( \dot{D}^{i_1 \ldots i_s}[\chi] + \gamma_5 \varepsilon^{i_1 jk} \pd_j D\indices{_k^{i_2 \ldots i_s}}[\chi] \right)\, .
\end{equation}
It is invariant under $SO(2)$ rotations
\begin{equation}
\chi \rightarrow e^{\alpha \gamma_5} \chi
\end{equation}
generated by $\gamma_5$, mixing the two chiral components of $\chi$. This extends the results of \cite{Bunster:2012jp} for spins $1/2$ and $3/2$ to arbitrary half-integer spin.

In this section, we prove that this is the action that one would
obtain starting from the usual Fang--Fronsdal action for $\psi$ and
solving the constraints. As in the bosonic case, the argument is
indirect and relies on the fact that the action is (almost) uniquely
determined by its invariance properties.

The Fang--Fronsdal action is given by
\begin{equation}\label{action_fermi_compact}
S =  -  i  \int\! d^4\! x\,  \bar{\psi}^{\,\mu_1 \cdots \mu_s} \, \mathcal{G}_{\mu_1 \cdots \mu_s} \, , 
\end{equation}
where the tensor $\mathcal{G}$ is given in terms of the Fronsdal tensor \eqref{eq:fronsdaltensor} by
\begin{equation}
  \mathcal{G}_{\mu_1 \cdots \mu_s}
  =
  \mathcal{F}_{\mu_1 \cdots \mu_s} - \frac{s}{2} \gamma_{(\mu_1} \bslashed{\mathcal{F}}_{\,\mu_2 \cdots \mu_s)} - \frac{s\left(s-1\right)}{4} \eta_{(\mu_1 \mu_2} \mathcal{F}^{\, \prime}_{\mu_3 \cdots \mu_s)} \nonumber
\end{equation}
and the field satisfies the triple gamma-trace condition
\begin{equation}
\bslashed{\psi}^{\ \prime}_{\,\mu_4 \cdots \mu_s} = 0 \, .
\end{equation}
This trace condition can obviously be solved to express everything in
terms of the field variables $\psi_{i_1 \cdots i_s}$,
$\psi_{0 i_2 \cdots i_s}$ and $\psi_{00 i_3 \cdots i_s}$ with at most
two temporal indices (see for example \cite{Campoleoni:2017vds} for
the explicit formulas). Equivalently, one can perform the invertible
change of variables \eqref{eq:xidef} to eliminate the components with
two zeros in favor of $\Xi_{i_3 \cdots i_s}$. Once this is done, the
equations of motion show that $\psi_{i_1 \cdots i_s}$ and
$\Xi_{i_3 \cdots i_s}$ are dynamical variables, while
$\psi_{0 i_2 \cdots i_s}$ is a Lagrange multiplier for the constraint
as already indicated above. Therefore, the action
\eqref{action_fermi_compact} necessarily takes the form
\begin{align}
S = \int \! dt \,d^3\!x\, \big[ \, &\Theta^A(\Psi_B) \dot{\Psi}_A  - \mathcal{H}(\Psi_A) \\
&+ \bar{\psi}_{0 i_2 \cdots i_s} \mathcal{C}^{i_2 \cdots i_s}(\Psi_A) + \bar{\mathcal{C}}_{i_2 \cdots i_s}(\Psi_A) \psi^{0 i_2 \cdots i_s} \, \big] \, .
\end{align}
Here, we wrote $\Psi_A = (\psi_{i_1 \cdots i_s}, \Xi_{i_3 \cdots i_s})$ for the dynamical variables. Important points are:
\begin{enumerate}
\item the variables $\psi_{0 i_2 \cdots i_s}$ only appear as Lagrange
  multipliers for the constraints $\mathcal{C}(\Psi_A) = 0$;
\item these constraints are equivalent to equation \eqref{const};
\item since the original action is of first order in derivatives, the
  functions $\Theta^A$ contain no derivatives, while $\mathcal{H}$
  contains one spatial derivative only.
\end{enumerate}
The explicit form of $\Theta^A$, $\mathcal{H}$ and $\mathcal{C}$
beyond the features mentioned above are not necessary for the purposes
of this argument.

Now, when the constraints are solved in terms of the prepotentials as
is the previous section, the Lagrange multipliers
$\psi_{0 i_2 \cdots i_s}$ disappear from the action. Then, the kinetic
term $\Theta^A(\Psi_B) \dot{\Psi}_A$ must be a function of the
prepotentials with one time derivative and $2s$ spatial derivatives
(since the fields $\Psi_A$ are expressed as $s$ spatial derivatives of
the prepotential $\chi$). Since it must be invariant under gauge and
Weyl transformations of the prepotential, it must (up to integration
by parts) take the form of the first term of the action
\eqref{eq:actionprepot}. Similarly, the Hamiltonian density must
contain $2s + 1$ derivatives and, by the same invariance argument,
must take the form of the second term of \eqref{eq:actionprepot}. The
relative factor of these two terms is fixed by the fact that this
action should yield \eqref{eq:eomprepot} as equations of motion with
the relative factor written there, since these equations are a
consequence of the original Fronsdal equations as we saw in the
previous sections.

\section{Comments and Conclusions}
\label{Conclusions}

In this paper, we have developed the (linearized) conformal geometry
of higher-spin fermionic fields in three dimensions. The difficulty
comes from the fact that the Weyl tensor identically vanishes in three
dimensions so that the conformal invariants must be constructed out of
the ``Cotton tensor'', which generalizes the tensor with the same name
of gravity and involves higher order derivatives. This tensor was
defined and its central properties (gauge completeness and conformal
Poincaré lemma) were established.

We then used these conformal tools to introduce the ``prepotentials'',
which provide the explicit solution of the constraint equations
resulting from the Fang--Fronsdal action \cite{Fang:1978wz}. The
reformulation in terms of the prepotentials, intimately connected with
the twisted self-duality reformulation (see Eq.\ \eqref{eq:eomprepot}),
puts on the same footing the fermionic fields and the bosonic fields,
for which a similar prepotential formulation was achieved in
\cite{Henneaux:2015cda,Henneaux:2016zlu} starting from the Fronsdal
action \cite{Fronsdal:1978rb}.

The prepotential formulation possesses two striking features:
\begin{itemize}
\item The prepotential action enjoys both generalized diffeomorphism
  invariance (like the higher-spin (Fang--)Fronsdal action) {\em and}
  generalized Weyl invariance. This is true for all spins and holds
  both in the bosonic and fermionic cases. Furthermore, the spatial
  dimension is the critical dimension where Weyl geometry requires the
  introduction of the Cotton tensor since the Weyl tensor identically
  vanishes. This is also a feature that appears to be universal and
  was found to hold in higher dimensions where the prepotentials have
  a non trivial Young mixed symmetry
  \cite{Bunster:2013oaa,Henneaux:2016opm,Henneaux:2017xsb,Henneaux:2018rub,Lekeu:2018kul}.
  The emergence of higher-spin Weyl invariance deserves further
  understanding.
\item The resulting prepotential action always takes the same simple
  form, for all spins, namely ``prepotentials $\times$ (time
  derivative of the Cotton tensors $+$ curl of the Cotton tensors)''
  (see also Table \ref{tab:summary} at the end of this work). This is
  suggestive that the sum over all spins of the actions should enjoy
  remarkable symmetry properties, in particular $sp(8)$-symmetry
  \cite{Fronsdal:1985pd,Gelfond:2015poa} or
  hypersymmetry~\cite{Aragone:1979hw,Berends:1979wu,Berends:1979kg,Aragone:1980rk,Aragone:1983sz}
  (for more recent work see
  \cite{Bunster:2014fca,Fuentealba:2015jma,Fuentealba:2015wza}).
\end{itemize}

Finally, it would be of interest to extend the analysis of this paper
to \mbox{(anti-)de} Sitter backgrounds, following the spin-2 case
\cite{Julia:2005ze,Boulanger:2018shp,Boulanger:2018adg}. In that
context, we point out reference \cite{Kuzenko:2018lru}, which gives in
its Appendix C a construction of higher-spin Cotton tensors for
generic conformally flat space in three dimensions. That interesting
construction proceeds along different lines from those followed in our
paper and starts from the Fang–-Fronsdal field strengths.

\section*{Acknowledgments} 
 
We thank Xavier Bekaert, Andrea Campoleoni and Sergio Hörtner for useful
discussions. V.L.\ and A.L.\ are Research Fellows at the Belgian
F.R.S.-FNRS. This work was partially supported by the ERC through the
``High-Spin-Grav'' Advanced Grant and by FNRS-Belgium (convention FRFC
PDR T.1025.14 and convention IISN 4.4503.15).
This project has received funding from the European Research Council
(ERC) under the European Union's Horizon 2020 research and innovation
programme (``Exceptional Quantum Gravity'', grant agreement No 740209).

\begin{landscape}
\begin{table}
  \centering
  \resizebox{\linewidth}{!}{
  \begin{tabular}{l | l l}
    \toprule
                                          & Spin $s$ bosons                                                                                                                                                                                               & Spin $s+\frac{1}{2}$ fermions                                                                                                                                                                                                                   \\ \midrule
\rowcolor{blue!10}
    Prepotentials                         & $Z^a_{i_1 \cdots i_s}$, $a=1,2$                                                                                                                                                                               & $\chi_{i_1 \cdots i_s}$                                                                                                                                                                                                                         \\ 
    Higher-spin diffeomorphisms $\xi$     & $\Gamma Z^a = s \partial \xi^a + \frac{s(s-1)}{2} \delta \lambda^a$                                                                                                                                           & $\Gamma \chi = s \partial \xi +  s \gamma\lambda$                                                                                                                                                                                               \\ 
    and Weyl transformations $\lambda$    & $\Gamma Z^a_{i_1 \cdots i_s} = s \partial_{(i_1} \xi^a_{i_2 \cdots i_s)} + \frac{s(s-1)}{2} \delta_{(i_1 i_2} \lambda^a_{i_3 \cdots i_s)}$                                                                    & $\Gamma \chi_{i_1 i_2 \cdots i_s}  = s \partial_{(i_1} \xi_{i_2 \cdots i_s)} +  s \gamma_{(i_1} \lambda_{i_2 \cdots i_s)}$                                                                                                                      \\
    \rowcolor{blue!10}
                                          & $G^{a} = \left(\varepsilon \cdot \partial \, \cdot \,\right)^s Z^{a}$                                                                                                                                         & $G= \left(\varepsilon \cdot \partial \, \cdot \,\right)^s \chi$                                                                                                                                          \\
    \rowcolor{blue!10}
    Einstein tensor $G$                   & $G^{a\, i_1 \cdots i_s} = \varepsilon^{i_1 j_1 k_1} \cdots \varepsilon^{i_{s} j_{s} k_{s}}  \partial_{j_1} \cdots \partial_{j_{s}} Z^{a}_{k_1 \cdots k_{s}}$ & $G^{i_1 \cdots i_s} = \varepsilon^{i_1 j_1 k_1} \cdots \varepsilon^{i_{s} j_{s} k_{s}}  \partial_{j_1} \cdots \partial_{j_{s}} \chi_{k_1 \cdots k_{s}}$ \\
    \rowcolor{blue!10}
                                          & $\Gamma G^a =  \frac{s(s-1)}{2} \left( - \pd^2 \mu^a[\lambda] + \delta \Delta \mu^a[\lambda] \right)$, \, $\pd \cdot G^a = 0$                                                                                 & $\Gamma G = s (\varepsilon \cdot \pd \cdot \gamma) \mu[\lambda]$, \, $\pd \cdot G = 0$                                                                                                                                                                                                   \\
    \multirow{5}{*}{Schouten tensor  $S$} & $S^{a} = \sum_{p=0}^{\lfloor s/2 \rfloor} c_p  \delta^p G_{[p]}^{a}$                                                                                                                                          & $S = \sum_{p = 0}^{\lfloor s/2 \rfloor} a_p  \delta^p G_{[p]} + \sum_{p = 0}^{\lfloor (s-1)/2 \rfloor} b_p \delta^p \gamma \slashed{G}_{[p]}$                            \\
                                          & $S^{a \, i_1 \cdots i_s} = \sum_{p=0}^{\lfloor s/2 \rfloor} c_p  \delta^{(i_1 i_2}  \cdots \delta^{i_{2p-1} i_{2p}} G_{[p]}^{a \, i_{2p+1}  \cdots i_s)}$                                                     & $S^{i_1 \cdots i_s} = \sum_{p = 0}^{\lfloor s/2 \rfloor} a_p  \delta^{(i_1 i_2}  \cdots \delta^{i_{2p-1} i_{2p}} G_{[p]}^{i_{2p+1}  \cdots i_s)}$                                         \\
                                          &                                                                                                                                                                                                               & $\quad + \sum_{p = 0}^{\lfloor (s-1)/2 \rfloor} b_p \delta^{(i_1 i_2}  \cdots  \delta^{i_{2p-1} i_{2p}} \gamma^{i_{2p+1}} \slashed{G}_{[p]}^{i_{2p+2}  \cdots i_{s-1})}$ \\
                                          & $c_p = \frac{(-1)^p}{4^p p!}\frac{s\, (s-p-1)!}{(s-2p)!}$                                                                                                                                                     & $a_p = \frac{(-1)^p}{4^p\,p!}\frac{\left(s-p\right)!}{\left(s-2p\right)!}$,\, $b_p  =- \frac{1}{2} \frac{(-1)^p}{4^p\,p!} \frac{\left(s-p-1\right)!}{\left(s-2p-1\right)!}$                                                                                                                         \\
                                          & $\Gamma S^a = - \frac{s(s-1)}{2} \pd^2 \nu^a[\lambda]$, \, $\pd \!\cdot\! S - (s-1) \pd \bar{S} = 0$                                                                                                          & $\Gamma S = \pd \nu[\lambda]$, \,                                       $\pd \!\cdot\! S - \slashed{\pd}\slashed{S} - (s-1) \pd \bar{S} = 0$                                                                                                                                                               \\
                                   
 \rowcolor{blue!10}                    & $D^{a} = \left(\varepsilon \cdot \partial \, \cdot \,\right)^{s-1} S^{a}$                                                                                                                           & $D=\left(\varepsilon \cdot \partial \, \cdot \,\right)^{s} S$                                                                                                                     \\
 \rowcolor{blue!10}  Cotton tensor $D$ & $D^{ai_1 \cdots i_s} = \varepsilon^{i_1 j_1 k_1} \cdots \varepsilon^{i_{s-1} j_{s-1} k_{s-1}}  \partial_{j_1}  \cdots \partial_{j_{s-1}} S^{a}_{k_1 \cdots k_{s-1}}{}^{i_s}$                        & $D^{i_{1}\cdots i_{s}}=\varepsilon^{i_1 j_1 k_1} \cdots \varepsilon^{i_{s} j_{s} k_{s}}  \partial_{j_1} \cdots \partial_{j_{s}} S_{k_1 \cdots k_{s}}$ \\
 \rowcolor{blue!10}                    & $\Gamma D^a = 0$, \, $\bar{D}^a = 0$, \, $\pd \cdot D^a = 0$                                                                                                                                              & $\Gamma D = 0$, \, $\slashed{D} = 0$, \, $\pd \cdot D = 0$                                                                                                                            \\
    Action                             & $S[Z]= \frac{1}{2}  \int \dd t \dd^3x \ Z^a_{i_1 \cdots i_s}\left( \varepsilon_{ab}\dot D^{b\, i_1 \cdots i_s}  -  \delta_{ab} {\varepsilon^{i_{1}j}}_{k} \pd_{j} D^{b\, k i_2 \cdots i_s} \right)$ & $S[\chi] = - i \int \! dt \,d^3\!x\,  \chi^\dagger_{i_1 \ldots i_s} \left( \dot{D}^{i_1 \ldots i_s} + \gamma_5 \varepsilon^{i_1 jk} \pd_j D\indices{_k^{i_2 \ldots i_s}} \right)$ \\
    \rowcolor{blue!10}
    Equations of motion                & $\varepsilon_{ab}\dot D^{b\, i_1 \cdots i_s}  -  \delta_{ab} {\varepsilon^{i_{1}j}}_{k} \pd_{j} D^{b\, k i_2 \cdots i_s} = 0$                                                                           & $\dot{D}^{i_1 \ldots i_s} + \gamma_5 \varepsilon^{i_1 jk} \pd_j D\indices{_k^{i_2 \ldots i_s}}=0$                                                                                                             \\
    $SO(2)$ duality symmetry           & $Z^a \mapsto R\indices{^{a}_{b}} Z^b$; $R \in SO(2)$                                                                                                                                                & $\chi \mapsto e^{\alpha \gamma_5} \chi$                                                                                                                                           \\\bottomrule
  \end{tabular}}
\caption{This table summarizes the most important quantities to
  formulate bosonic and fermionic fields in $(3+1)$ dimensions in the
  prepotential formalism and their important properties. In order to
  overcome any notational ambiguity that might arise we provide the
  definitions in the index-free and the index notation. The bosonic
  formulation is taken from \cite{Henneaux:2015cda,Henneaux:2016zlu},
  and the fermionic formulas are taken from this work.}
  \label{tab:summary}
\end{table}
\end{landscape}

\appendix

\section{Proof of \texorpdfstring{$\gamma q = p$}{gamma q = p}}
\label{sec:proof-gamma-q}

We want to prove that equation \eqref{eq:EqForQ} of Section
\ref{sec:gauge-completeness}
\begin{equation}
\gamma^{rs} q_{i_1 \cdots i_{s-2} r \, j_1 \cdots j_n s} = p_{i_1 \cdots i_{s-2} \, j_1 \cdots j_n}
\end{equation}
always possesses a solution for arbitrarily given $p$'s with $(s-2,n)$
Young symmetry (the $q$'s have $(s-1, n+1)$ Young symmetry). It is a
linear system of inhomogeneous equations
\begin{align}
  \label{eq:inhsys}
  {A_{\alpha}}^A\, q_A = p_\alpha
\end{align}
that has a solution for given $p_\alpha$ if and only if the given
$p_\alpha$ fulfills $y^\alpha\, p_\alpha = 0$ for any left eigenvector
$y^\alpha$ of the matrix ${A_{\alpha}}^A$ with eigenvalue zero (i.e.,
$y^\alpha {A_{\alpha}}^A = 0$). Since we want no restriction on
$p_\alpha$, the matrix ${A_{\alpha}}^A$ should have no left
eigenvector for the eigenvalue zero. That is, the system
${A_{\alpha}}^A\, q_A = p_\alpha$ has always a solution for arbitrary
$p$'s if and only if the only solution to the equations
$y^\alpha {A_{\alpha}}^A = 0$ is $y^\alpha = 0$. In our case
\eqref{eq:inhsys} explicitly corresponds to
\begin{align}
  {A_{i_1 \cdots i_{s-2} \, j_1 \cdots j_n}}^{i'_1 \cdots i'_{s-2} r' \, j'_1 \cdots j'_n s'}\, q_{i'_1 \cdots i'_{s-2} r' \, j'_1 \cdots j'_n s'}
  = p_{i_1 \cdots i_{s-2} \, j_1 \cdots j_n}
\end{align}
where
\begin{align}
  {A_{i_1 \cdots i_{s-2} \, j_1 \cdots j_n}}^{i'_1 \cdots i'_{s-2} r' \, j'_1 \cdots j'_n s'}
  = \mathbb{P} \left( \delta^{i'_{1}}_{i_{1}} \cdots \delta^{i'_{s-2}}_{i_{s-2}}
  \gamma^{r' s'}
  \delta^{j'_{1}}_{j_{1}} \cdots \delta^{j'_{n}}_{j_{n}} \right)
\end{align}
and $\mathbb{P}$ projects on the appropriate Young symmetries, i.e., $(s-2,n)$ for the lower indices and $(s-1, n+1)$ for the higher indices.
Transposing spinor indices, this is equivalent to showing that the
only solution to the equation
\begin{equation}
\label{eq:gamy}
 \gamma_{(r|(s} y_{i_1 \cdots i_{s-2}) | j_1 j_2\cdots j_n)} = 0
\end{equation}
with $ y_{i_1 \cdots i_{s-2} \, j_1 j_2\cdots j_n}$ of $p$-symmetry
type $(s-2,n)$ is $y_{i_1 \cdots i_{s-2} \, j_1 j_2\cdots j_n} = 0$.

As a first step we set $i_{k}=s$ and $j_{k}=r$ which leads for
$r \neq s$, for which $\gamma_{rs}$ is invertible, to
$ \gamma_{rs} y_{s \cdots s r \cdots r} = 0$ which implies
\begin{align}
  \label{eq:ssrr}
y_{s \cdots s r \cdots r} =0 \,.
\end{align}
In this section we do no sum over the $s$ and $r$ indices. We now use
\eqref{eq:ssrr} to ``free'' one of the indices of the first group so
that
$\gamma_{(r|(s} y_{i_1 s \cdots s) | r \cdots r)} = \gamma_{r(s}
y_{i_1 s \cdots s) r \cdots r}= \# \gamma_{rs} y_{i_1 s \cdots s r
  \cdots r}=0$, which leads for $r \neq s$ to
$\gamma_{rs} y_{i_1 s \cdots s r \cdots r} = 0$ and therefore
\begin{align}
  \label{eq:onefree}
y_{i_{1}s \cdots s r \cdots r} =0 \, .
\end{align}
We denote with $\#$ strictly positive constants whose values are not
relevant for the proof. Repeating this argument for
$\gamma_{(r|(s} y_{i_1 i_{2} s \cdots s) | r \cdots r)} = 0$ implies
that $y_{i_{1} i_{2} s \cdots s r \cdots r} =0$. We can reiterate on
the first group of indices and repeat the same analysis for the second
group to get
\begin{align}
   y_{i_1 \cdots i_{s-2}  r \cdots r} = 0 = y_{s \cdots s j_{1} \cdots j_{n}} \,.
\end{align}
We next use these relations to connect the two indices groups since
$\gamma_{(r|(s} y_{i_1 s\cdots s) | j_1 r \cdots r)} = \# \gamma_{rs}
y_{i_1 s\cdots s j_1 r \cdots r}=0$ which leads for $r \neq s$ to
$y_{i_1 s\cdots s j_1 r \cdots r} =0$. Now we can systematically
``free'' the indices, e.g.,
$\gamma_{(r|(s} y_{i_1 i_{2} s\cdots s) | j_1 r \cdots r)} =0$ implies
$y_{i_1 i_{2} s\cdots s j_1 r \cdots r}=0$. Repeating this analysis
iteratively on both groups of indices shows that the only solution of
\eqref{eq:gamy} is $y_{i_1 \cdots i_{s-2} \, j_1 j_2\cdots j_n} = 0$
which completes the proof.

\bibliographystyle{utphys} 
\bibliography{bibl} 

\end{document}